# Nonlinear dynamics of human locomotion: effects of rhythmic auditory cueing on local dynamic stability

*Running title:* Auditory cueing and gait stability

*Authors:* Philippe Terrier[1,2*], Olivier Dériaz[1,2]

*Affiliations:*
1. IRR, Institute for research in rehabilitation, Sion, Switzerland
2. CRR, Clinique romande de réadaptation SuvaCare, Sion, Switzerland

*Correspondence:*
[*]Philippe Terrier
Clinique romande de réadaptation
Recherche médicale
Av. Gd-Champsec 90
1951 Sion
SWITZERLAND
Tél.  +41(0)27 603 20 70

E-mail : philippe.terrier@crr-suva.ch

**Abstract**

Synchronizing steps with an external auditory stimulus (rhythmic auditory cueing (RAC) enhances gait recovery in neurological disorders. The activation of specific sensory-motor processes, which may partially replace impaired neural pathways, is likely the cause of the observed benefits. Nonlinear indexes, such as scaling exponents and Lyapunov exponents, have been proposed to characterize RAC effects. The maximal Lyapunov exponent, which measures how fast a non-linear system diverges from initial perturbation, estimates the degree of resilience of gait control to small perturbations, i.e. the local dynamic stability (LDS). The objective of the present study was to assess to what extent RAC influences gait LDS, and to compare this effect with that on scaling exponents. Twenty healthy subjects performed 6x5min walking trials on an instrumented treadmill at three different speeds (slow, normal, fast). Freely chosen walking cadences were measured during the first three trials and then imposed accordingly in the last three trials with a metronome. The 2D trajectory of the center of pressure on the treadmill was recorded. From the antero-posterior and the medio-lateral signals, both long-term and short-term LDS were computed. Long-term LDS was strongly enhanced (relative change +47%), with significant change in every direction and speed. The average change in short-term LDS was smaller (+3%), with a more marked effect at low speed (+5%). RAC substantially modified the fluctuation dynamics of the center of the pressure trajectory. We also observed that both LDS and fractal dynamics (scaling exponents) responded similarly to RAC. Thus, both scaling exponents and LDS are responsive to sensory-motor synchronizing processes that RAC activates, and may constitute relevant indexes for evaluating gait variability in cued walking. Finally, the more locally stable gait pattern could be an indication of a lower fall risk, which may be an advantage to patients of RAC therapies.

**Introduction**
During walking, individuals are able to voluntarily adjust their gait to external cues, such as floor markers, metronomes, or the moving belt of a motorized treadmill. External spatial or temporal stimuli could facilitate movement: namely, cued walking exhibits positive effects on various gait characteristics of neurologically impaired patients (Thaut and Abiru, 2010), such as patients with Parkinson's Disease (PD) (Nieuwboer et al., 2007), or stroke (Thaut et al., 2007;Roerdink et al., 2009). In PD patients, synchronizing steps with an external rhythmic stimulus (Rhythmic Auditory Cueing, RAC), significantly improves walking speed, stride length and cadence (Lim et al., 2005). Similarly, it has been suggested that a treadmill could act as an external cue to enhance gait rhythmicity and reduce gait variability (speed cueing) (Frenkel-Toledo et al., 2005). In PD patients, it has been recently observed that treadmill training further improved gait capacity compared to standard (overground) gait training (Bello et al., 2013). In stroke patients, treadmill exercise has been shown to improve walking velocity (Pohl et al., 2002;Luft et al., 2008), with a concomitant increased activation in the cerebellum and midbrain, as evidenced by fMRI (Luft et al., 2008;Enzinger et al., 2009). Combining several cues together seems to provide further enhancements: in PD patients, treadmill training associated with auditory and visual cues might give better results than more conventional treatments (Frazzitta et al., 2009).

Although empirical evidence supports the use of cued walking in neurorehabilitation practice, many aspects of the underlying neurophysiological mechanisms are not yet fully understood (Bello and Fernandez-Del-Olmo, 2012). It is thought that motor control adjusts gait movement in order to provide stride by stride an optimal combination of step length and step time, which results in steady walking speed. This optimal combination coincides with minimal energy expenditure (Zarrugh et al., 1974;Donelan et al., 2001). Furthermore, due to the inherent instability of bipedal locomotion, a manageable range of kinematic fluctuations is crucial, because pushing the center of mass too far away from an optimal trajectory could lead to a fall. Therefore, continuous control of stride-to-stride fluctuations is required in order to minimize energy expenditure and fall risks. Those continuous optimizations likely imply both feedforward (from internal models) and feedback (from sensory inputs) mechanisms (Kuo, 2002), which require low attentional demands and are highly automated: the existence of specific structures at the spinal level (central pattern generators) is strongly suspected (Dimitrijevic et al., 2006). In contrast, synchronizing movement with rhythmic auditory cues requires complex supraspinal mechanisms, which induce increased neuronal activity in sensorimotor cortex, supplementary motor area, premotor cortex, inferior parietal cortex, basal ganglia and cerebellum (Repp and Su, 2013).

While the neurophysiological mechanisms of cued walking have not yet been fully characterized, it is evident that synchronizing gait with external stimuli activates specific supraspinal/cortical processes, which add to basic gait control. A sign of this activation can be found in the modification of stride-to-stride fluctuation patterns. Indeed, an interesting feature of gait control is that reasonable deviations from the mean persist across subsequent strides: time series of basic gait parameters (Stride Time, ST; Stride Length, SL; Stride Speed, SS) exhibit substantial auto-correlation or statistical persistence, probably due to feedback loops in gait control (Hausdorff et al., 1995;Terrier et al., 2005). In other words, a larger stride as compared to average SL is more likely to be followed by subsequent larger strides. Statistical persistence can be assessed by means of scaling exponents using detrended fluctuation analysis (DFA)(Hausdorff et al., 1996;Terrier and Dériaz, 2012). In overground walking, synchronizing the gait with a metronome substantially modifies the stride-to-stride fluctuation pattern: a strong anti-persistence in ST has been observed (Terrier et al., 2005;Sejdic et al., 2012), while a persistent pattern is conserved in SL and SS (Terrier et al., 2005). Anti-persistence means that deviations in one direction are statistically more likely to be followed by subsequent deviations in the opposite direction (i.e., longer strides are more likely to be followed by shorter strides). Similarly, treadmill walking (speed cueing) seems to induce an anti-persistent pattern in SS only (Dingwell et al., 2010), while ST and SL remain persistent. When both treadmill and RAC are combined, all three parameters (ST, SL, and SS) are anti-persistent (Terrier and Dériaz, 2012). The anti-persistent pattern could be induced by fast over-correction of deviations in the controlled variable, which results in continuous oscillations around target values (Dingwell et al., 2010;Terrier and Dériaz, 2012).

Because cued walking requires specific supraspinal/cortical control mechanisms (Zijlstra et al., 1995;Dingwell et al., 2010;Terrier and Dériaz, 2012), one could wonder whether those activations divert motor control from performing gait stabilization tasks. It is well known that dividing attention between gait and a cognitive task may impair gait stability (dual tasks paradigm) (Woollacott and Shumway-Cook, 2002;Weerdesteyn et al., 2003). It has been recently observed that treadmill walking and RAC increased reaction time (Peper et al., 2012):  attentional demands were further elevated during visually and auditively cued walking as compared to uncued walking. However, opposite to classical dual-task situations, during cued walking, the attention is not divided between gait and another cognitive task; on the contrary, the increased attentional demands are devoted to a specific gait control task. Therefore, it could be assumed that the increased control over the gait would lead to a higher level of stability and thus to lower falling risk, which could benefit patients with gait disorders. When considering whether to submit patients to complex multi-cued rehabilitation protocols (Frazzitta et al., 2009), it is important to first better know the stability outcome of cueing tasks in healthy individuals.

The faculty to maintain steady progression despite the constant presence of small internal control errors or small external disturbances is referred to as Local Dynamic Stability (LDS). Gait LDS can be characterized using the maximal Lyapunov exponent, which is a parameter that assesses how fast the system diverges (Brown, 1996;Dingwell and Cusumano, 2000;Dingwell, 2006;Terrier and Deriaz, 2011). With this method (Rosenstein et al., 1993), local divergence exponents ($\lambda$) are computed from the slopes of divergence curves, which quantify how fast the neighboring trajectories of a reconstructed state space diverge from nearest neighbor points. Strictly speaking, due to the nonlinearity of the divergence curves, multiple slopes could be defined. Thus, no true single maximal Lyapunov exponent exists. Different slopes quantify local divergence (and thus local stability) at different timescales. Classically, two different timescales have been proposed to assess gait LDS: long-term divergence $\lambda_l$ (long-term LDS (Dingwell and Cusumano, 2000)) and short-term divergence $\lambda_s$ (short-term LDS (Dingwell et al., 2001)). The former is based on the time interval between four and 10 strides after the initial perturbation, and the latter is based over a time interval corresponding to one stride (or one step) after the initial perturbation.

Although there is no theoretical reason that directly associates LDS and falling risk, if the rate of divergence is small (high LDS), it is likely a sign of tighter control over the gait. When an unexpected perturbation occurs, a walking individual must execute appropriate avoidance strategies to not fall. It can be assumed that, if motor control can efficiently manage small perturbations (high LDS), it can also cope with large perturbations that would lead to falling. Modeling studies (Roos and Dingwell, 2010;Bruijn et al., 2011), as well as experimental studies in healthy individuals (McAndrew et al., 2011;van Schooten et al., 2011;Hak et al., 2012), provide some evidence that short-term LDS is actually related to fall risk. Furthermore, recent clinical studies observe that older people at risk for falling exhibited lower short-term LDS (Lockhart and Liu, 2008;Toebes et al., 2012). Long-term LDS has been suspected to be responsive to certain conditions (Chang et al., 2010;McAndrew et al., 2011;Sloot et al., 2011) and diseases (Dingwell et al., 2000). An enhanced long-term LDS could be related to compensatory mechanisms under destabilizing situations (van Schooten et al., 2011) or to a more cautious gait (Chang et al., 2010). While it is still uncertain whether LDS could constitute a relevant and usable proxy for fall risk, LDS is a valid non-linear measure of gait variability based on sound theoretical background (Dingwell, 2006;Stergiou and Decker, 2011). It can therefore serve to highlight potential changes in the fluctuations of continuously measured kinematics variables induced by various conditions, such as cued walking. It has been shown that the treadmill induces higher LDS as compared with overground walking (Dingwell et al., 2001;Terrier and Deriaz, 2011). Similarly, it has been recently observed that RAC induced a higher long-term LDS in overground walking (Sejdic et al., 2012). Although there is already some evidence that cued walking has a stabilizing effect on gait, there is still a need to characterize the effects of the combination of both treadmill and RAC.

The statistical persistence/anti-persistence and LDS of walking characterize distinct aspects of gait control (Terrier and Deriaz, 2011). Assuming an autoregressive stochastic process, statistical persistence quantifies temporal dynamics of discrete events (ST, SL, SS) over hundreds of consecutive strides, and serves to characterize the feedbacks in locomotor control (Dingwell et al., 2010;Terrier and Deriaz, 2011;Terrier and Dériaz, 2012). Assuming chaotic system, LDS quantifies temporal dynamics in continuous signals (acceleration, speed, position) and assesses the degree of resilience of motor control to small perturbations over shorter timescales (Dingwell, 2006;Roos and Dingwell, 2010). Although these two indexes seem loosely related from a theoretical point of view, they are both responsive to cued walking (Dingwell et al., 2001;Terrier and Deriaz, 2011;Sejdic et al., 2012). A treadmill experiment revealed a strong correlation between long-term LDS and statistical persistence (Jordan et al., 2009), but this was contradicted by others (Terrier and Deriaz, 2011). Therefore, more results are needed to assess whether correlations exist between LDS and statistical persistence and to clarify the mechanism behind such a potential association.

The objective of the present study was to analyze the effect of RAC on LDS in healthy, middle-aged individuals walking on an instrumented treadmill, which measured the trajectory of the center of pressure. Both short-term and long-term LDS were assessed. The hypothesis was that the increased control over the gait while combining both treadmill and RAC results in a more locally stable gait. In addition, the correlation between LDS and statistical persistence (data from a companion article (Terrier and Dériaz, 2012)) was evaluated. The overall aim is to determine whether LDS could constitute a relevant index for studying cued walking.

**Methods**
The present study is based on raw data obtained in a previous study (Terrier and Dériaz, 2012). Please refer to this article for further information about the experimental procedure.

**Participants**
Twenty healthy subjects (10 females, 10 males) took part in the study. The participants' characteristics were (mean (SD): age 36 yrs (11), body mass 71kg (15), and height 171cm (9). The experimental procedure was approved by the local ethics committee (Commission Cantonale Valaisanne d'Ethique Médicale, Sion, Switzerland).

**Experimental procedure**
The testing sessions consisted of two series of three 5min.30 treadmill walking: 30s of habituation to the speed, and 5min of measurement. Treadmill speeds imposed on the subjects were: Preferred Walking Speed (PWS), 0.7 x PWS (low speed) and 1.3 x PWS (high speed). The speed sequence was randomly attributed. The trials with the "metronome" condition (Treadmill+RAC) were performed at the same speeds as the first 3 trials. The imposed cadences were the preferred cadences, which were measured during the first trials without metronome.

The measurement device was a motorized treadmill (FDM-TDL, Scheinworks/Zebris, Schein, Germany), instrumented with foot-pressure sensors aimed at dynamic plantar pressure assessment (100Hz sampling rate, 128x56 pressure sensors on a 108.4 x 47.4cm grid). A ''movie'' of the feet pressure on the treadmill belt was obtained (as illustration, see online supplementary materials, video S1). The raw data consisted for each trial of 30,000 frames of 7,168 points. They were exported for subsequent analysis with Matlab (Mathworks, MA, USA). Complementary statistical analysis was realized with Stata (StataCorp, TX, USA).

*Data analysis*
The continuous trajectory of the center of pressure was computed as the weighted average of the pressure data, and using the standard method for determining the barycenter (Sum of mass x position)/(Sum of mass). The two axes of the trajectory consisted of an anteroposterior (AP) component (along the direction of the displacement of the treadmill belt) and a mediolateral component (ML, perpendicular to the displacement). Figure 1 presents a typical plot of center of pressure trajectory, with the corresponding AP and ML signals. The raw trajectory can be also seen in the video provided in the online supplementary material.

The raw 100Hz signals were filtered down to 50Hz in order accelerate the subsequent steps of data analysis; an eighth-order low pass Chebyshev Type I filter was used, which filtered the signal in both the forward and reverse directions to remove all phase distortion (Matlab command *decimate*). Step Frequency (SF), and thus average step duration, was assessed by calculating the Fast Fourier Transform of the AP signal. Then, a duration corresponding to 175 strides was selected from the raw signals. The resulting segments, whose length depended upon the SF of each participant at each speed condition, were time-standardized to a uniform length of 10,000 samples, by using a polyphase filter implementation (Malab command *resample*).

The method for quantifying LDS has been described in many articles (Dingwell and Cusumano, 2000;Lockhart and Liu, 2008;Terrier and Deriaz, 2011). More theoretical information is provided in the appendix at the end of the article. In short, the method is based on the examination of characteristics of a time series that is embedded in an appropriately constructed state space, which contains a sufficient number of independent coordinates to define the state of the system unequivocally (Takens, 1980). The state space was reconstructed according to the Takens' theorem, as classically applied in gait dynamics studies (Dingwell and Cusumano, 2000). The time delay and the embedding dimension were assessed by the average mutual information (AMI) function and global false nearest neighbors (GFNN) analysis, respectively. A time delay of 15 and 18 samples, respectively, was used for the ML and AP directions. A constant dimension of six was set for all the directions. These values corresponded to the average results of the AMI and GFNN analyses. The Lyapunov exponent is the mean exponential rate of divergence of initially nearby points in the reconstructed state space, as defined in (Rosenstein et al., 1993) for small data sets. Then, time was normalized by the average stride time of each trial, taking into account the resample to 10,000 samples. As in other studies (Dingwell and Cusumano, 2000;Dingwell et al., 2001;Yakhdani et al., 2010;van Schooten et al., 2011), two divergence exponents were computed: short-term LDS over the timescale corresponding to the first step ($\lambda_s$) and long-term LDS ($\lambda_l$) over the timescale between the 4$^{th}$ and 10$^{th}$ strides. In order to illustrate the influence of RAC on the divergence dynamics, the mean logarithmic divergence curve was computed by averaging each sample across participants (N=20). In addition, Standard Deviation (SD) was computed at seven discrete points (Figure 2).

*Statistics*

We analyzed four dependent variables: (1) short term LDS in the antero-posterior direction ($\lambda_s$-AP); (2) short term LDS in the medio-lateral direction ($\lambda_s$-ML); (3) long term LDS in the antero-posterior direction ($\lambda_l$-AP); (4) long term LDS in the medio-lateral direction ($\lambda_l$-ML). The independent variables were speed (3 level) and cueing condition (treadmill and treadmill+RAC), but in the present study we were interested mainly in the effect of RAC. The descriptive statistics consisted of the mean and standard deviation for each condition (fig. 3-4). In addition, the spread of the individual results were presented by using notched boxplots (median and quartiles, N=20 participants, Figure 3 and 4). Standardized Effect Size (ES=delta(mean)/SD$_{pooled}$, i.e. Hedges's g) was computed in order to describe the strength of the effect of RAC (Cohen, 1992;Nakagawa and Cuthill, 2007). The precision on the effect sizes was estimated with 95% Confidence Intervals (CI). CI were ±1.96 times the asymptotic estimates of the standard error of g. Graphical representations of ES and corresponding CI are shown for each variable and speed condition (Figure 3 and 4). Arbitrary thresholds for medium (0.5), large (0.8) and huge (2) effects (Cohen, 1992) were used in order to ease the interpretation. It should be reminded that the analysis of ES and CI is strictly equivalent to the paired-t test. Furthermore, in order to minimize type I error risk induced by the multiple comparisons (3 different speeds, 2 directions), the analysis was completed using a multivariate comparison test (Hotelling's T-squared test) separately for long-term and short-term LDS, which is similar to omnibus ANOVA testing: The null hypothesis H0 was that the mean differences (Treadmill+RAC minus Treadmill) were equal to zero.

Potential associations between the LDS and the statistical persistence were assessed using Canonical Correlations Analysis (CCA), after removing potential speed effects. The main advantage of CCA is that it reduces the risk of type I errors that increase when multiple correlations are performed. Persistence data were obtained from the results of the previous study (Terrier and Dériaz, 2012), in which detrended fluctuation analysis (DFA) was used to characterize statistical persistence using scaling exponents (α). First, a linear regression was computed between the average speed and each dependent variable, separately for each cueing condition (N=60). The spread of the average speeds among individuals can be found in another companion article (Terrier, 2012). Then, the residuals of the linear regression were computed by subtracting the predicted values from the data (observed response minus predicted response). The residuals reflect the remaining variance when linear speed effects are removed. The CCA is a multivariate statistical method that assesses the strength of association between two sets of variables (Anderson et al., 2006). The relationship (canonical function) between two linear composites (variates) is computed. The canonical correlation coefficient expresses the strength of the relationship between the two variates that compose the canonical function. Three sets of variables were defined for each condition: from the results of the present study, set#1: [$\lambda_s$-AP; $\lambda_s$-ML], set#2 [$\lambda_l$-AP; $\lambda_l$-ML]; from the results of the previous study, set #3 [α-ST; α-SL; α-SS]. Two CCAs were realized for each condition,

set#1 vs. set#3 and set#2 vs. set#3. The significance of the canonical correlations (i.e. r <> 0) was assessed with the Wilks' lambda statistics. Furthermore, the analysis was completed with redundancy results, which express the amount of variance in one set explained by the linear composite (canonical variate) of the other set.

**Results**
Figure 2 shows the average logarithmic divergence <ln [$d_j(i)$]> in both cueing conditions (i.e., treadmill and treadmill+RAC). A very different divergence regime is observed. While very few differences are evident over the first stride (short term LDS), the curve reaches a plateau faster under the treadmill + RAC condition.

The descriptive statistics for the short term LDS ($\lambda_s$) are shown in the upper panels of Figure 3. The results of the multivariate $T^2$ test revealed that a significant effect of RAC is likely ($T^2=34$, $p=0.01$), but the average relative change is small (-3.0%). A lower λ (lower divergence rate) signifies that the LDS was higher. The partial ES results (Fig. 3, lower panels) are contrasted, but it seems that a relevant effect is effective at lower speeds ($\lambda_s$-AP: -5.6%, ES: -1.0; $\lambda_s$-ML: -4.5%, ES: -0.9). At PWS, the results are $\lambda_s$-AP: -3.0%, ES: -0.5; $\lambda_s$-ML: -1.2%, ES: -0.2. At fast speeds, the results are $\lambda_s$-AP: -1.1%, ES: -0.3; $\lambda_s$-ML: -2.3%, ES: -0.4.

The descriptive statistics for the long term LDS ($\lambda_l$) are shown in the upper panels of Figure 4. The results of the multivariate $T^2$ test shows a highly significant effect of RAC ($T^2=301$, $p<0.0001$), with a large average relative change (-47.0%). The partial ES results (fig. 4, lower panels) revealed large effects (lower λ => higher LDS) across all directions and speeds: slow speed: $\lambda_l$-AP: -49%, ES: -2.8; $\lambda_l$-ML: -56%, ES: -2.9; PWS: $\lambda_l$-AP: -44%, ES: -3.3; $\lambda_l$-ML: -48%, ES: -2.5; fast speed: $\lambda_l$-AP: -43%, ES: -2.4; $\lambda_l$-ML: -42%, ES: -2.1.

Table 1 summarizes the results of the CCA. The two canonical orthogonal functions resulted in two canonical correlation coefficients, which are presented in the first and second columns, with corresponding p-values in the third and forth columns. Only the redundancy results of the first canonical function are shown in the last two columns. Regarding results for the short-term LDS (set#1 vs. set#3), the hypothesis that a correlation exists with statistical persistence should be rejected. Indeed, low (0.05-0.30), not significant, correlation coefficients are observed and the redundancy results show that the canonical functions explain a very small part of the variance in the other set (2% to 4%). On the contrary, a relevant association between the long-term LDS and the statistical persistence is very likely, especially under the treadmill+RAC condition ($p<0.001$). In addition, the redundancy results of the first canonical function reveal that a large part of the variance in one canonical variate is explained by the other canonical variate (25%-74%).

**Discussion**
By measuring the trajectory of the center of pressure on a motorized treadmill, the objective of the present study was to analyze the responsiveness of Local Dynamic Stability (LDS) to Rhythmic Auditory Cueing (RAC) in healthy individuals. RAC slightly increased short-term LDS, with an effect that was especially evident at slow speeds. On the other hand, a huge effect of RAC on long-term LDS was observed: LDS was largely improved for all speeds and directions. Correlation results revealed that a relevant association between long-term LDS and statistical persistence (scaling exponent α) is likely, especially under the "treadmill + RAC" condition. On the contrary, an association between statistical persistence and short-term LDS is very unlikely.

*Methodological considerations*
As far as we know, the present study proposes for the first time computing LDS from the trajectory of the center of pressure obtained from an instrumented treadmill. As illustrated in Figure 1 and in the online movie (supplementary material), small deviations in the trajectory is evident from one stride to the next, which are likely linked to the continuous adjustments that the motor control performs to maintain stable gait. Other authors have used the center of pressure in gait stability studies (Day et al., 2012), for instance to analyze how motor control reacts to large external perturbations (Hof et al., 2010). The center of pressure trajectory seems therefore a relevant parameter, from which LDS can be computed. Moreover, the results of the present study are comparable to those of a recent study that analyzed the response of LDS to RAC in overground walking (Sejdic et al., 2012), which supports the fact that the method was adequate.

As in other recent LDS studies (Yakhdani et al., 2010;van Schooten et al., 2011;McAndrew Young and Dingwell, 2012), this study used a normalized sample size (10,000) and a normalized number of strides (175). It also employed uniform time delays and dimensions. As proposed by others (Yakhdani et al., 2010;van Schooten et al., 2011), this study computed short-term LDS over one step, and not one stride. Regarding short-term LDS, the total variance (Figure 3 and 4), which is the combination of the actual biological inter-individual variability, the actual intra-individual variability and the measurement error, was rather low: expressed as CV (SD/mean), it lies between 6.1% and 7.0%. Regarding long-term LDS, which is known to exhibit lower intra-session repeatability (Kang and Dingwell, 2006), CV lies between 11.8% and 23.6%. In comparison, in a previous treadmill study that used trunk accelerometry and a less standardized methodology (Terrier and Deriaz, 2011), we observed an average CV of 21% for short-term LDS and 39% for long-term LDS. Because both studies included the same number of subjects who were sampled from the same population, the difference is very likely the measurement error. Consequently, the combination of standardized procedures and use of the center of pressure trajectory probably makes it possible to obtain a lower measurement error and thus higher reliability, which increases the statistical power and reduces the risk of type II errors.

*Effects of rhythmic auditory cueing*
The logarithmic divergence curves, such as presented in Figure 2, were strikingly modified by RAC. It is known that with the Rosenstein's algorithm a plateau is reached in the divergence curve when no divergence can further occurs because of the limits of the attractor (Figure A2). In other words, the trajectories form a flow that is bounded. As hypothesized in the introduction, RAC enabled specific sensory-motor synchronizing processes, which added to automated gait control: that additional control probably narrowed the maximal bounds in the state space. The probable modification of the attractor bounds is logically also reflected in long-term LDS results, because it is computed from the slope close to the plateau (Figure 2 and Figure A2). Thus, long-term LDS was strongly enhanced (lower $\lambda_l$, ES>2, relative change 47%). On the other hand the change in short-term LDS was smaller (ES -0.55, 3% relative change). Because it measure the immediate response to small perturbations, short-term LDS is probably more appropriate to evaluate actual gait stability and fall risk (Roos and Dingwell, 2010;Bruijn et al., 2013), and thus is a more relevant parameter for clinical applications (Toebes et al., 2012). Because patients with neurological gait disorders tend to walk slowly, it is worth noting that the effect of RAC on short-term LDS was stronger at slow speeds (Fig. 3), for which a relative change of about 5% was measured. On the other hand, treadmill walking improves short-term LDS by 9% as compared with overground walking (Terrier and Deriaz, 2011). Provided that short-term LDS is a relevant proxy for fall risk, combining both treadmill and RAC for slow walking patients seems therefore a quite safe intervention.

*Comparative responsiveness of short-term and long-term local dynamic stability*
The results suggest that a specific modality of gait control (i.e. the synchronization with an external cue) may affect differentially short-term and long-term LDS through a modification of divergence curves.

Three theoretical studies based on artificial gait modeling attempted to better understand the relationships between $\lambda_s$, $\lambda_l$, and actual fall risk. With a 2-D passive model, Su and Dingwell (Su and Dingwell, 2007) showed that short-term stability $\lambda_s$ increased linearly with the mean amplitude of applied perturbations, but not $\lambda_l$, which remained unchanged. With an improved 3-D active model, they subsequently showed that $\lambda_s$ was responsive to noise amplitude applied to the lateral step controller, while $\lambda_l$ was not responsive (Roos and Dingwell, 2010). Interestingly, contrary to human results, $\lambda_l$ was around zero, which may indicate an attractor with narrow limits: the authors explained that "the noise applied to the controller was dampened out quickly" (Roos and Dingwell, 2010). An independent study, based on 2-D passive modeling and using alternative methods to induce perturbations to the gait model, confirmed that $\lambda_s$ relates to the probability of falling(Bruijn et al., 2011). Moreover, they observed only a weak relationship between $\lambda_l$ and actual stability.

Two recent human studies further analyzed the use of LDS as an index for global stability and falling risk by inducing perceptual perturbations to healthy individuals. Van Schooten et al. (van Schooten et al., 2011) used galvanic vestibular stimulation to impair balance. They confirmed that $\lambda_s$ could be used to assess global stability of gait. However, they reported that the impaired balance decreased $\lambda_l$ (improved stability). They explained that that "may be due to compensatory changes, which occur at longer timescales [...]". The same contradictory stability outcome has been described in (McAndrew et al., 2011): by inducing visual and mechanical perturbations to healthy individuals, they observed increased $\lambda_s$ and decreased $\lambda_l$. They showed divergence curves shifting up and to the left under destabilizing conditions, with a steeper slope in the short term (higher $\lambda_s$), and then a flatter (lower $\lambda_l$), and higher plateau in the long term. The authors explained that the divergence curves reached their maximum local divergence limits more quickly during perturbed walking.

Taking into consideration this short review of the literature and the results of the present study, we propose that a parallel should be made between the fractal-like, persistent fluctuation patterns that are observed among consecutive strides (Terrier et al., 2005;Dingwell and Cusumano, 2010;Terrier and Deriaz, 2011) and the positive long-term LDS. In other words, as motor control allows deviations from the mean to persist across strides ($\alpha>0.5$), this translates into long-term local instability (positive $\lambda_l$). On the contrary, when motor control tightly regulates gait parameters, for instance by attempting to synchronize with RAC, the persistent pattern is replaced by oscillations around the target value (anti-persistence (Terrier and Dériaz, 2012)), more stationary time series take place (Terrier, 2012), and the local divergence is more quickly dampened within an attractor exhibiting narrower bounds (Figure 2 and 4), as in the gait models (Roos and Dingwell, 2010). On the other hand, we hypothesize that short-term LDS $\lambda_s$ is more related to rapid automated/unconscious motor processes that constantly hinder uncontrolled growth of small perturbations and manage obstacle avoidance (Weerdesteyn et al., 2004). That would explain why $\lambda_s$ is a relevant proxy for fall risk (Roos and Dingwell, 2010;Bruijn et al., 2011;McAndrew et al., 2011;van Schooten et al., 2011;Bruijn et al., 2013). The opposite response of $\lambda_s$ and $\lambda_l$ (McAndrew et al., 2011;Sloot et al., 2011) is therefore more likely due to compensatory mechanisms: by altering rapid feedback mechanisms, perceptual perturbations induce not only lower short-term LDS (higher $\lambda_s$), but also a more cautious, voluntary controlled gait, which results in higher long-term LDS (lower $\lambda_l$), as induced by RAC.

*Correlations between local dynamic stability and statistical persistence*
The results of the CCA revealed that a relevant correlation exists between statistical persistence and long-term LDS. This means that individuals that presented a more persistent pattern in stride-to-stride fluctuation tend to also have lower long-term LDS (higher $\lambda_l$). The relationship is stronger under the treadmill+RAC condition. In this case, individuals that presented a more anti-persistent pattern in stride-to-stride fluctuation (lower α) tended to have higher long-term LDS (lower $\lambda_l$). A previous study (Terrier and Deriaz, 2011) also showed a moderate correlation between α-ST and $\lambda_l$ (r=0.28 and 0.42), which was not significant due to a smaller sample size. An independent study also observed a strong correlation between long-term LDS and statistical persistence during treadmill walking (r=0.72) (Jordan et al., 2009). Overall, these results reinforce the hypothesis that both long-term instability (positive $\lambda_l$) and statistical persistence/anti-persistence are the manifestation of a common underlying motor control process. In contrast, $\lambda_s$ seemed not correlated with statistical persistence (Table 1), which corroborates the hypothesis that short-term LDS is related to an independent motor control process.

**Conclusion**
There is conclusive evidence that synchronizing gait to external clues substantially modifies the stride-to-stride fluctuation dynamics (Terrier et al., 2005;Dingwell and Cusumano, 2010;Sejdic et al., 2012;Terrier and Dériaz, 2012), the stationarity of gait parameters (Terrier, 2012) and the long-term LDS (Figure 1 and 3 and (Sejdic et al., 2012)). It is very likely that those substantial modifications are induced by the activation of specific cortical sensory-motor synchronization mechanisms (Halsband et al., 1993;Zijlstra et al., 1995;Egerton et al., 2011), which partially replace (or add to) the automated regulation of the gait. This activation could be one of the underlying mechanisms that explains the benefits of cued walking in patients with neurological disorders. As a result, the abovementioned parameters should be assessed in order to evaluate neurological gait disorders and the outcome of cued walking intervention. In particular, an enhanced long-term LDS could indicate a more cautious gait (compensatory mechanism), as well as the presence of a less correlated pattern (lower α) in stride-to-stride fluctuations (Herman et al., 2005).

The main practical outcome of the present study is that combining RAC and treadmill does not impair gait stability in healthy individuals (Figure 1-3). Similarly, cued walking (RAC, as well as treadmill) seems to have no substantial effect on gait variability (Dingwell et al., 2001;Terrier et al., 2005;Terrier and Deriaz, 2011) or on step width (Terrier, 2012), two parameters that have also been used to assess gait instability and fall risk (Hausdorff et al., 2001;Owings and Grabiner, 2004;Nordin et al., 2010). Concerning neurological disorders, it has been shown that a combination of attentional strategy (focusing on big steps) and RAC reduced gait variability in PD patients (Baker et al., 2008), which corroborates with the hypothesis that cued walking redirects higher cognitive functions to gait, and thus compensates for automated gait regulation deficit. Similarly, a recent study attempted to analyze this issue by assessing the influence of RAC on obstacle avoidance capabilities in PD patients (Nanhoe-Mahabier et al., 2012). The experimental design combined treadmill walking and RAC, as in the present study. The authors observed that PD patients were able to successfully execute an obstacle avoidance task, when auditory cueing is administered simultaneously. They concluded: "our data suggest that PD patients can benefit from auditory cueing even under complex, attention-demanding circumstances, and that the metronome does not act as a dual task that negatively affects gait."

Although further studies are needed to analyze LDS and other variability indexes in various neurological gait disorders, the present study introduced new evidence that cued walking could be a valuable and safe treatment in gait rehabilitation.

**Conflict of interest**
None declared.

**Acknowledgments**
The authors thank M. Philippe Kaesermann for the loan of the instrumented treadmill. The study was supported by the Swiss accident insurance company SUVA, which is an independent, non-profit company under public law. The IRR (Institute for Research in Rehabilitation) is supported by the State of Valais and the City of Sion.

**Appendix**

The objective of this appendix is to summarize the theoretical background behind the assessment of the local dynamic stability (LDS) based on the Rosenstein's algorithm. Most of the concepts described below are adapted from (Rosenstein et al., 1993), (Fraser and Swinney, 1986), and (Kennel et al., 1992).

*Lyapunov exponents*
The Lyapunov exponents assess the sensitivity to initial conditions of a dynamical system, which is characterized by a finite number of $n$ state variables and $n$ equations. If an $n$-dimensional sphere of initial conditions is defined, the exponential rates of divergence are given by a spectrum of $n$ Lyapunov exponents ($\lambda$). These exponents describe a multidimensional ellipsoid expending/contracting with time along particular axes (Lyapunov directions), which are dependent upon the system flow. A particular Lyapunov exponent represents the local instability in a given direction. If the exponent is positive, it diagnoses chaos. When the system is globally stable the rate of contractions in some directions must counterbalance the rate of expansions in others in order to obtain a stable attractor. Thus, the sum across the entire spectrum of the Lyapunov exponents is negative. The largest (maximal) Lyapunov exponent ($\lambda_1 \geq \lambda_2 \geq \ldots \geq \lambda_n$) defines the direction along which the system is the most unstable, because the exponential growth in this direction will dominate growth along the other directions. It is defined by the following equation:

$$d(t) = Ce^{\lambda_1 t} \tag{A1}$$

$d(t)$ being the average divergence at time $t$ and C a constant that normalize the initial separation.

*Phase space reconstruction: Embedding*
When working with real world data, the set of $n$ equations characterizing a dynamical system is not available. However, the attractor dynamics can be reconstructed from a single time series according to the Takens's theorem (Takens, 1980). The process that unfolds the time series into a multidimensional state space is referred to as embedding. The reconstruction is based on the time delay method. The state of the system $X$ defined by a $N$-point time series $\{x_1, x_2 \ldots x_n\}$ at discrete time $i$ is:

$$X_i = (x_i \ \ x_{i+J} \ \ \ldots \ \ x_{i+(m-1)J}) \tag{A2}$$

Where $J$ is the lag or the reconstruction (or time) delay and $m$ is the embedding dimension.

In principle, for an infinite noise-free time series, the time delay $J$ and the embedding dimension $m$ can be arbitrarily chosen. In smaller data set of noisy data, a too small $J$ would produce a compressed attractor along the identity line (high correlation or excessive redundancy); conversely, a too large $J$ would produce causally disconnected attractor (irrelevance issue) (Fraser and Swinney, 1986;Kim et al., 1999). A solution is to choose a delay $J$ that maximizes the independence between the coordinates (for instance $x_i$ and $x_{i+J}$). A linear independence is realized when the autocorrelation function of the time series first pass through zero. However, non-linear dependence is likely in most dynamical systems. The method of the average mutual information (AMI) (Fraser and Swinney, 1986) solves this issue by assessing the general dependence of two variables (namely, $x_i$ and $x_{i+J}$) using the Shannon's information theory. *J* is determined from the first minimum in the mutual information function of the time series. Thus, $x_{i+J}$ adds the largest amount of information available to the first coordinate $x_i$ and conserves some correlation among them, which has been empirically proven to be a good equilibrium between redundancy and irrelevance (Fraser and Swinney, 1986).

The issue of the choice of a correct embedding dimension is more topological than dynamical. A phase space with a sufficient number of dimensions allows the attractor to deploy without self-crossing, thus correctly capturing the dynamic flow of the system. However, an excessive number of dimensions make difficult the manipulation of the reconstructed attractor (high computational time). Furthermore no dynamics is operating in the dimensions in excess and they are mostly populated by noise. Conversely, a too small number of dimensions increases the risk of spurious vicinity between points in the phase space due to projection effects. In other words, the distances between points are distorted. Thus, a too compact attractor contains a high number of these false neighbors, which makes difficult to account for the true flow of the system. The method of global false nearest neighbors (GFNN) assesses the rate of spurious vicinity that occurs in under-dimensioned phase spaces (Kennel et al., 1992). It differentiates the points that are neighbor due to the dynamics of the system (following the flow) from the points that are neighbor solely due to projection issues. The percentage of false neighbors is computed for increasing dimensions, and the minimal dimension for a correct embedding is found when the percentage is close to zero.

*Maximal Lyapunov exponent*
The rate of exponential divergence in the appropriately embedded attractor is computed as follows: First, nearest neighbor of each point on the attractor is located. The nearest neighbor $X_{\hat{j}}$ is found by searching the point that minimizes the distance to the particular reference point $X_j$:

$$d_j(0) = \min \|X_j - X_{\hat{j}}\| \tag{A3}$$

where $d_j(0)$ is the initial distance from the *j*th point to its nearest neighbor, and $\| \ \|$ denotes the Euclidian norm. An additional constraint is imposed, namely that the nearest neighbors must be separated by a time, which exceeds the mean period of the time series. As a result, the two neighbors are on a different orbit of the attractor. Then the Euclidian distances $d_j(i)$ between the two trajectories defined by the subsequent points *i* downstream the reference point and its nearest neighbor are computed, for instance as follows for the first iteration:

$$d_j(1) = \|X_{(j+1)} - X_{(\hat{j}+1)}\| \tag{A4}$$

It is expected that the rate of divergence between the trajectories is given by the equation (A1). In its logarithm form, the equation (A1) becomes:

$$\ln d_j(i) \approx \ln C_j + \lambda_1 (i\Delta t) \tag{A5}$$

where $C_j$ is the initial separation. This linear equation defines a set of parallel lines with slope equal to the maximal Lyapunov exponent. Thus, the average logarithmic divergence for all *j* is computed <ln *d_j(i)*>=*avg[d_1(i), d_2(i), ..., d_n(i)]*. The average logarithmic divergence <ln *d_j(i)*> can be represented in a diagram as a function of time (see Fig. 2 in the main text). Finally, the Lyapunov exponent is approximated using least-square fit to the average line:

$$y(i) = \frac{1}{\Delta t} \left\langle \ln d_j(i) \right\rangle \tag{A6}$$

*Divergence curves and application to gait dynamics*
Using the example of the Lorenz attractor, Rosenstein et al. described that plotting the average divergence as a function of time (Figure A2) showed "a long linear region" after a "short transition". A plateau occurred at longer times ("saturation") because the attractor is bounded in phase space, which implies that the average divergence cannot continuously grow with time. Thus, the linear fit to the divergence must be performed on "prominent" linear region. The authors acknowledge that the determination of that region is mostly empirical and may be difficult in real systems.

   Gait dynamics produces a divergence curve that does not exhibit a clear linear region (Figure A2). Whether gait could be considered as a real chaotic process described by a limited set of equations is therefore debatable. However, a pragmatic approach is to consider that the linear fitting estimates the local divergence over a given time scale, which is used to characterize the dynamics of the system, even if a true maximal Lyapunov exponent does not exist (Dingwell, 2006). Thus, the term "divergence exponent" may be more appropriate than the term "maximal Lyapunov exponent". Each time the foot is on the ground, it gives to the motor control the opportunity to thwart perturbations; consequently, the rate of divergence per stride (or per step) is more physiological than the rate of divergence per second (Bruijn et al., 2013). The long-term divergence fitted against a time scale between the fourth and the tenth stride has been adopted on a purely empirical basis (Dingwell and Cusumano, 2000). On the other hand, the short-term divergence fitted over the first stride (or first step) estimates the local dynamic stability immediately after a perturbation, which is more physiologically sound.

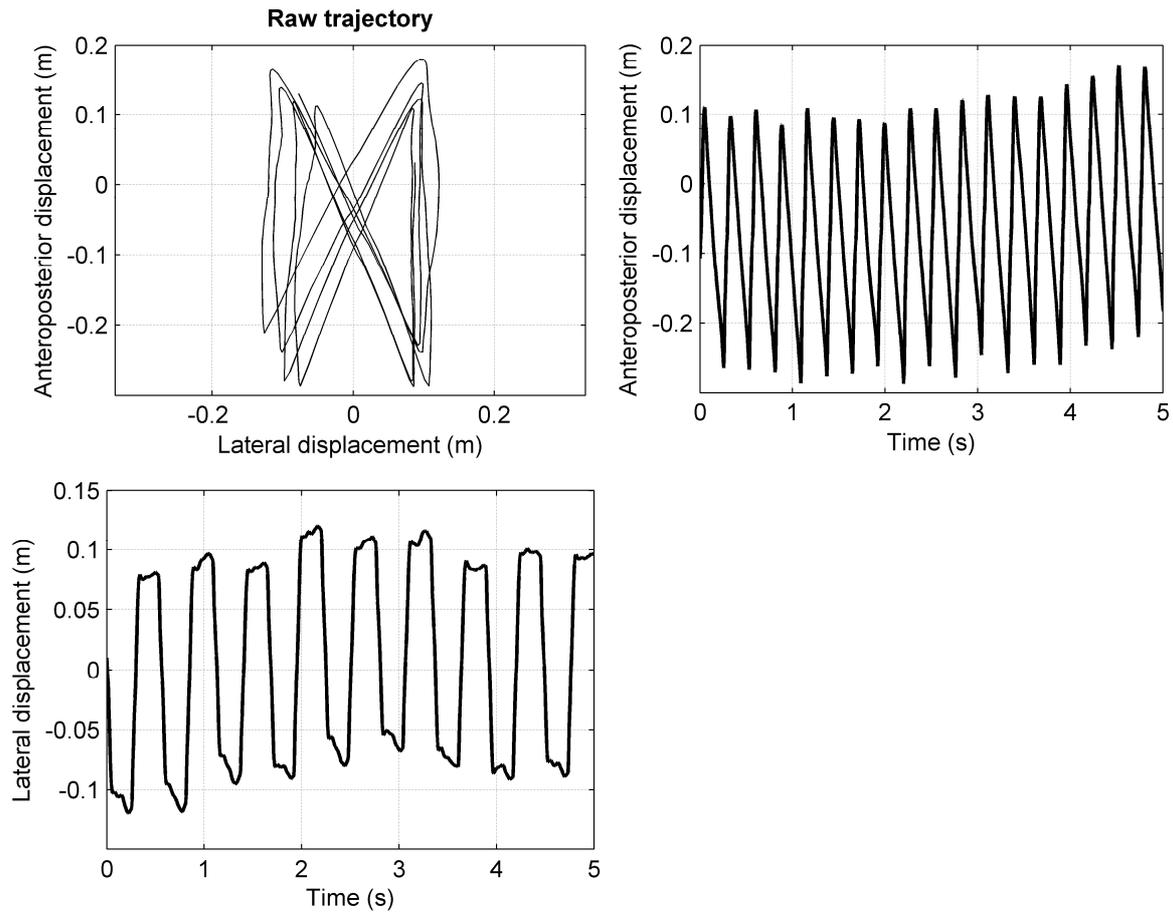

**Figure 1. Trajectory of the center of pressure.** A participant walking at preferred walking speed (1.25 m·s$^{-1}$) during 5min. on an instrumented treadmill, which measured the feet pressure on the treadmill belt. The trajectory of the center of pressure was computed (barycenter method). Only 5 seconds are shown. The upper-left panel shows the raw trajectory. The upper-right and lower panels show the decomposition of the raw trajectory in, respectively, the anteroposterior and the mediolateral directions.

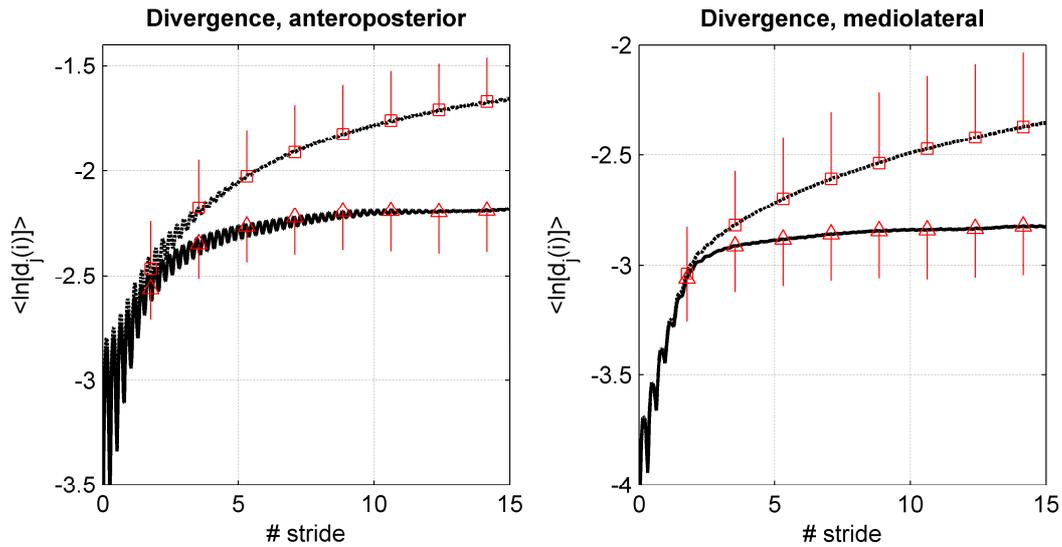

**Figure 2. Divergence diagrams**. The average logarithmic divergence ($<\ln[d_j(i)]>$) in antero-posterior and medio-lateral direction was measured in the reconstructed state space of the center of pressure trajectory (50Hz sampling rate), in 20 individuals walking at preferred walking speed. 175 consecutive strides were analyzed, normalized at 10,000 samples. The value at each time (50Hz) was averaged across the subjects (N=20). Time was normalized by the average stride time (1.14 s). Discontinuous lines (squares) are the results for the treadmill only condition. Continuous lines (triangles) are the results for the dual cueing condition (Treadmill + Rhythmic Auditory Cueing). Mean value at 100, 200, 300, 400, 500, 600, 700 and 800 samples are shown (squares and triangles) with the corresponding SD (vertical lines, N=20).

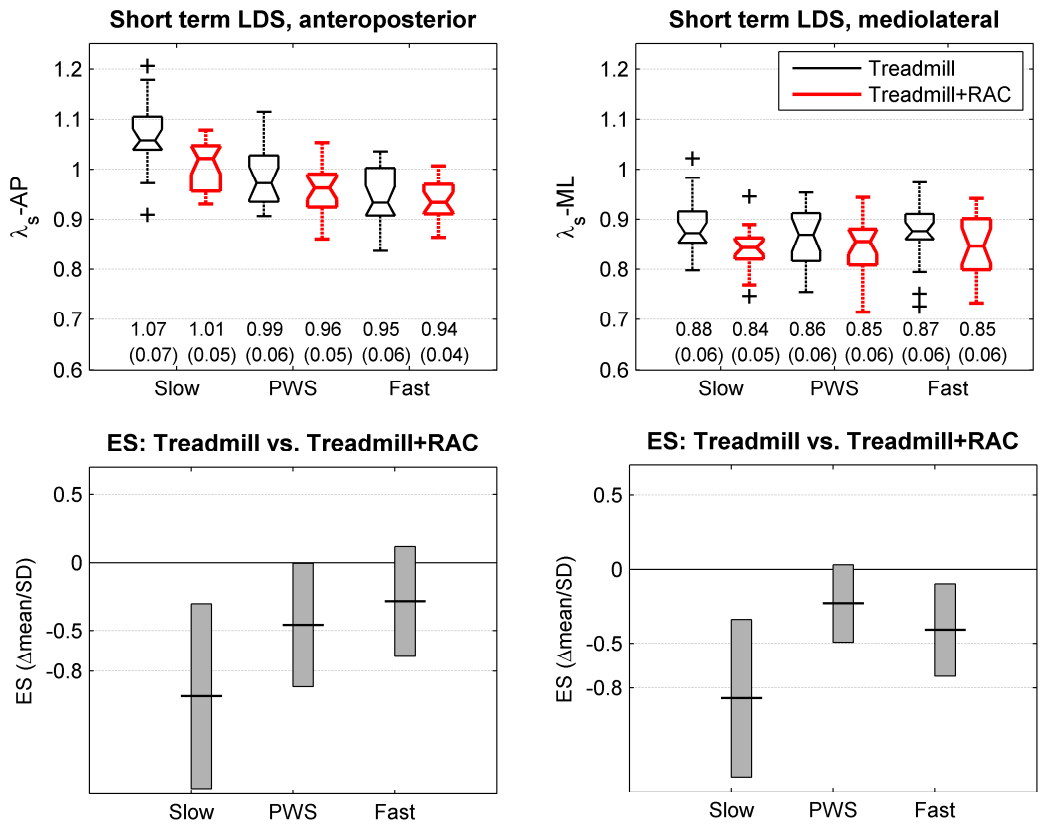

**Figure 3. Short-term local dynamic stability (LDS).** Twenty healthy subjects walked 3x5min. on an instrumented treadmill without (thin lines, black) and with Rhythmic Auditory Cueing (RAC, (metronome), thick lines, red) at their preferred cadence for the given speed. The center of pressure trajectory over 175 consecutive strides was analyzed along the anteroposterior and mediolateral axes. Short-term LDS is computed from the rate of logarithmic divergence over one stride (finite time Lyapunov exponents, $\lambda_s$). Selected speeds were Preferred Walking Speed (middle, PWS), 0.7 x PWS (left, Slow) and 1.3 x PWS (right, Fast). The range of individual results (N=20) is presented with notched boxplots. Printed values are mean(SD). Bottom panels show the effect size (ES) of the auditory cueing (i.e., the mean difference normalized by SD (Hedges's g)). Vertical boxes are the 95% confidence intervals for the effect size estimations.

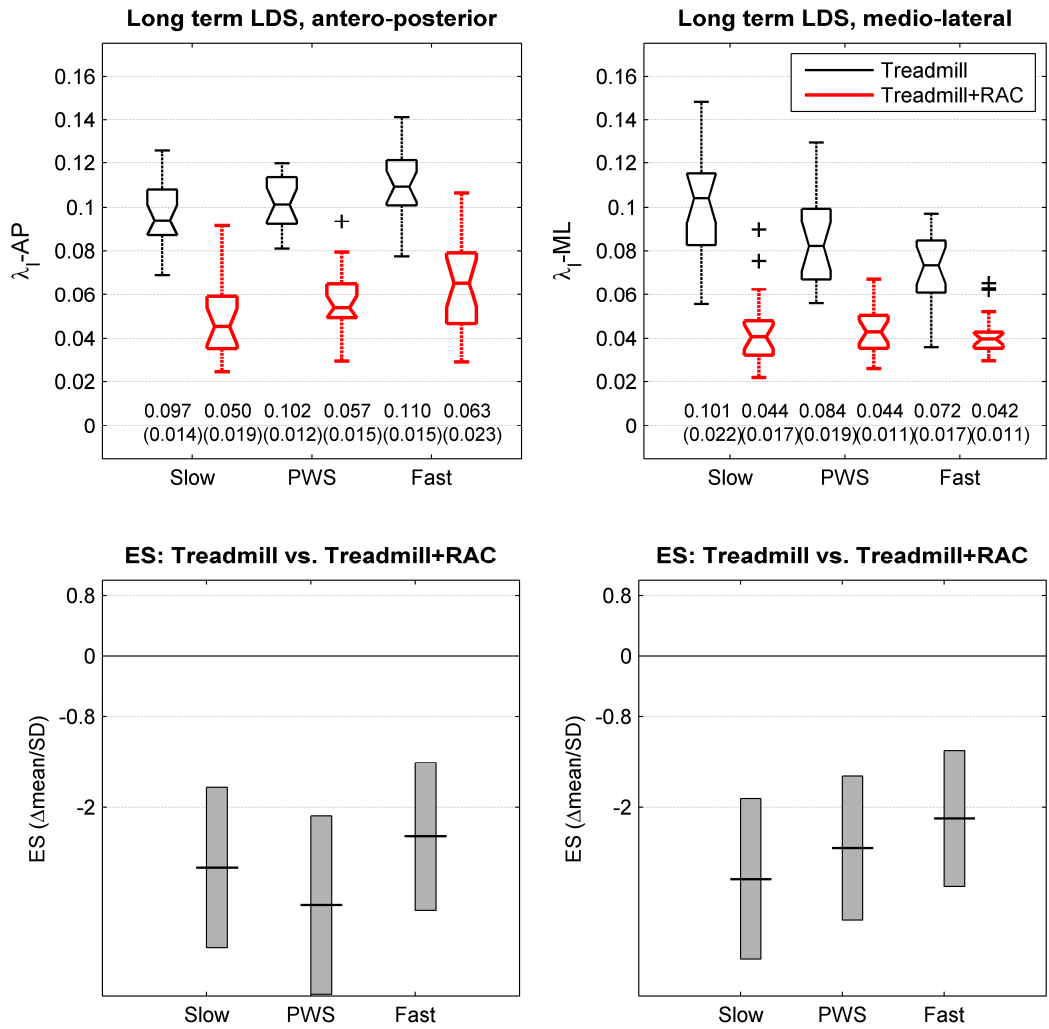

**Figure 4. Long-term local dynamic stability (LDS)**. Twenty healthy subjects walked 3x5min.on an instrumented treadmill without (thin lines) and with Rhythmic Auditory Cueing (RAC, (metronome), thick lines) at their preferred cadence for the given speed. The center of pressure trajectory over 175 consecutive strides was analyzed along the anteroposterior and mediolateral axes. Long-term LDS is computed from the rate of logarithmic divergence among the 4$^{th}$ to 10$^{th}$ consecutive strides (finite time Lyapunov exponents, $\lambda_l$). Selected speeds were Preferred Walking Speed (middle, PWS), 0.7 x PWS (left, Slow) and 1.3 x PWS (right, Fast). The range of individual results (N=20) is presented with notched boxplots. Printed values are mean(SD). Bottom panels show the effect size (ES) of the auditory cueing (i.e., the mean difference normalized by SD (Hedges's g)). Vertical boxes are the 95% confidence intervals for the effect size estimations.

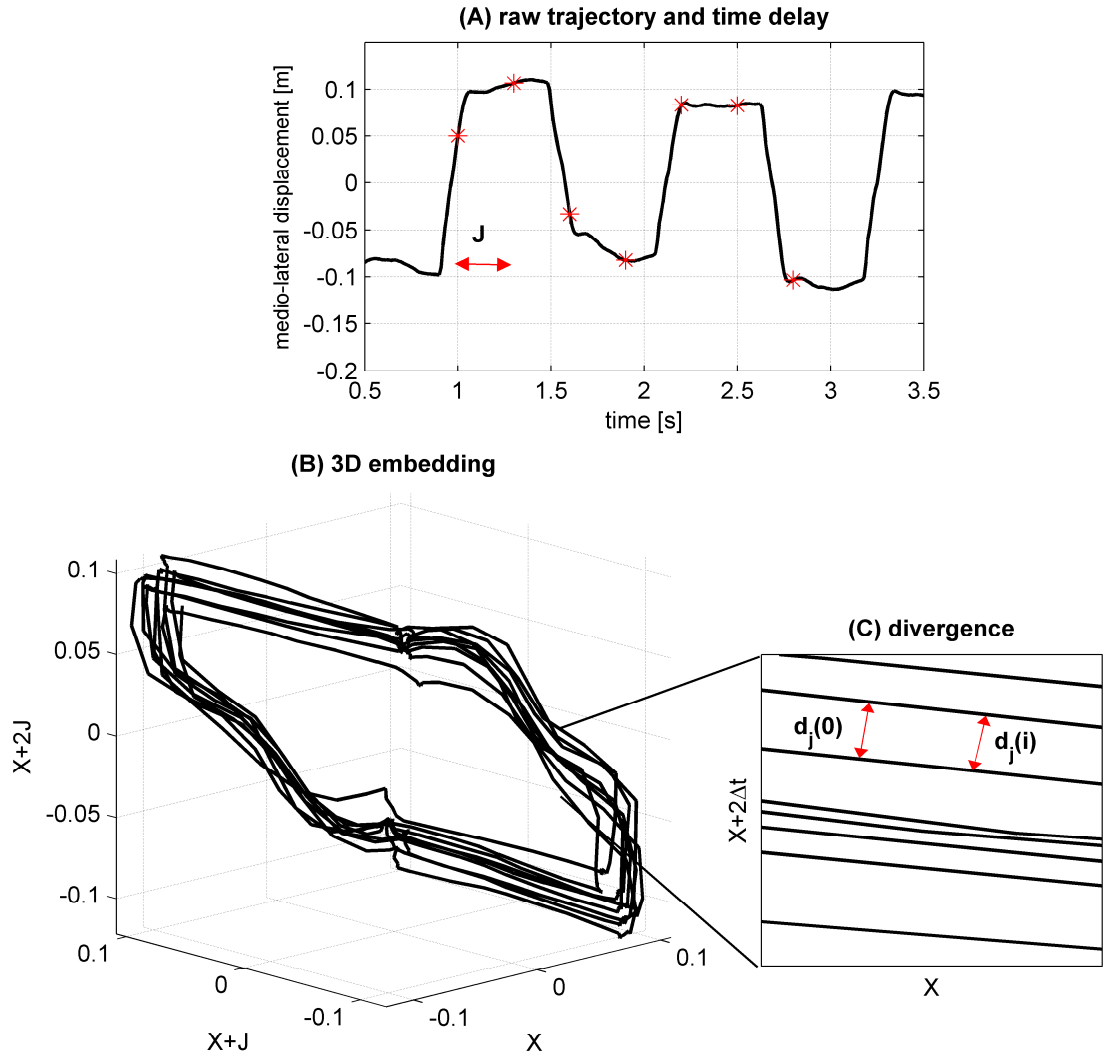

**Figure A1: Illustration of the computing of LDS with gait data.** Panel (A) shoes 3.5 second of the medio-lateral displacement of the center of pressure sampled at 50Hz. The result of the average mutual information analysis was 18 samples (0.36sec), which defined the reconstruction delay J. The results of the false neighbor analysis revealed that the minimal embedding dimension should be 6. Therefore, five time-delayed copies of the original time series have been built, according to the equation (2). Panel (B) displays the reconstructed attractor with J=18 and m=6. Only the first three dimensions are used to build a 3D projection of the 6D attractor. Panel (C) is a 2D magnification of the attractor. It shows the flow of the trajectories along which the maximal Lyapunov exponent will be computed. Two nearest neighbors (equation (A3)) are shown, separated by the initial distance $d_j(0)$. Two downstream points are separated by the distance $d_j(i)$ (equation (A4)).

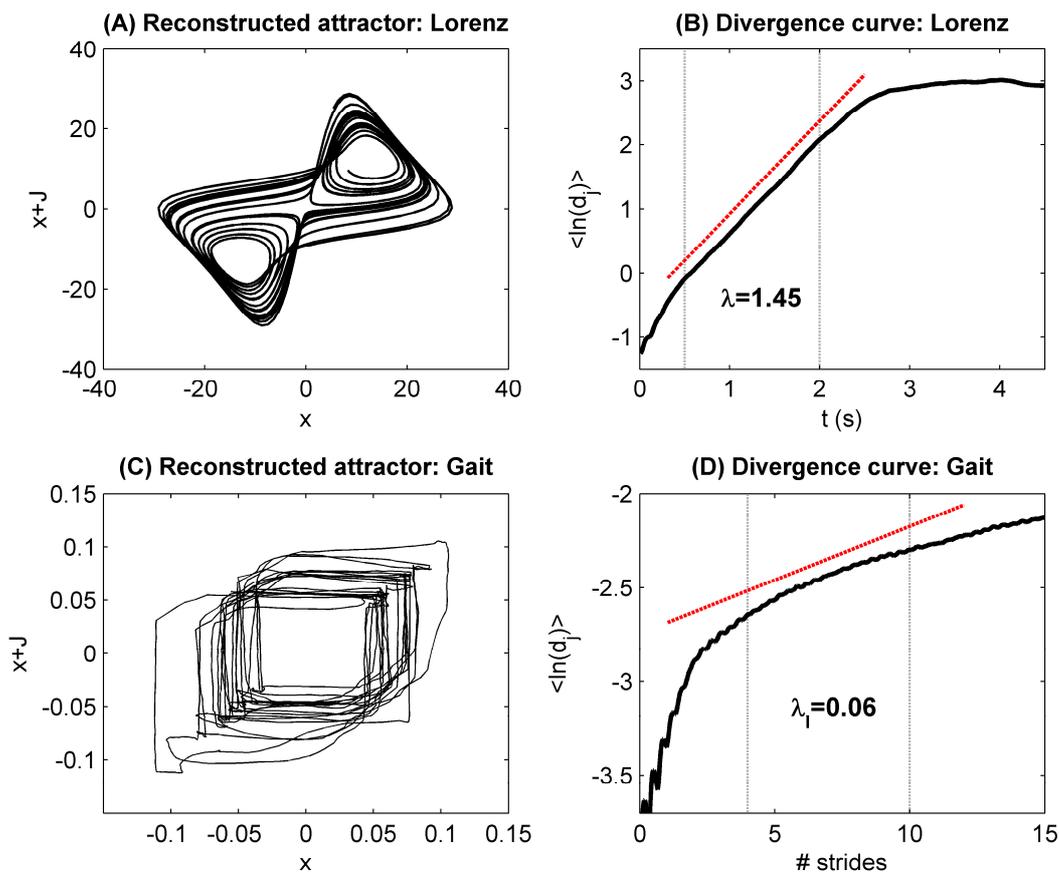

**Figure A2: Divergence curves.** Panel (A) shows the reconstructed Lorenz attractor build with the same parameters as in (Rosenstein et al., 1993). The corresponding divergence curve (i.e. the logarithmic average divergence as a function of time, equation (A6)) is presented in the panel (B). The "prominent" linear region between 0.5 s and 2 s is highlighted by the vertical dotted lines. The linear fit to the curve in that region is highlighted by the red dotted line, and the slope corresponds to the maximal Lyapunov exponent (equation (A5) and (A6)), $\lambda=1.45$. The true Lyapunov exponent for the Lorenz attractor is 1.5. The reconstructed attractor for gait data (mediolateral signal, Figure 1 and A1) is displayed in panel (C). The panel (D) shows the corresponding divergence curve, with time normalized by the average stride duration. The time scale for the computation of the long-term LDS (4-10 strides) is shown with the vertical dotted lines. The linear fit is displayed (red dotted line), with the slope value (divergence exponent $\lambda_l$).

**Supplementary online material**

**Movie M1. Illustration of the measurement of the centre of pressure.** Five seconds of analysis are shown, slowed down to 25 seconds. The pressure on the grid of sensors is shown with color ranging from blue (no pressure) to red (high pressure): the succession of 10 steps is visible. The dark red circle indicates the current position of the center of pressure. The dark red trace show the corresponding trajectory, which is also displayed in Figure 1.

**Table**

**Table 1. Canonical correlation analysis (CCA).**
Scaling exponents (α) evaluate the statistical persistence present in the time series of stride time (ST), stride length (SL) and stride speed (SS). The local dynamic stability assesses the local divergence (λ) of the trajectory of the center of pressure in anteroposterior (AP) and mediolateral (ML) directions. Two CCAs were performed for each cueing condition: 1) between short-term LDS (two variables, $\lambda_s$-AP and $\lambda_s$-ML) and scaling exponents (3 variables, α-ST, α-SL and α-SS); 2) between long-term LDS (two variables, $\lambda_s$-AP and $\lambda_s$-ML) and scaling exponents (3 variables, α-ST, α-SL and α-SS). All the seven variables were normalized by walking speed prior to CCA (N=60, i.e., 3 imposed speed times 20 participants for each variable). See text for more precision on the CCA method.

| | | Canonical correlation coefficients | | p values (Wilks' lambda statistics) | | Redundancy (first canonical function) | |
|---|---|---|---|---|---|---|---|
| Treadmill only | Short-term LDS vs. Scaling exponents | 0.30 | 0.09 | 0.44 | 0.81 | 2.9% | 4.0% |
| | Long-term LDS vs. Scaling exponents | **0.63** | 0.22 | 0.00 | 0.24 | 25.0% | 29.0% |
| Treadmill + RAC | Short-term LDS vs. Scaling exponents | 0.29 | 0.05 | 0.52 | 0.93 | 2.0% | 4.0% |
| | Long-term LDS vs. Scaling exponents | **0.91** | **0.56** | 0.00 | 0.00 | 51.0% | 74.0% |

26                                                                    Auditory cueing and gait stabilityEnzinger, C., Dawes, H., Johansen-Berg, H., Wade, D., Bogdanovic, M., Collett, J., Guy, C., Kischka, U., Ropele, S., and Fazekas, F. (2009). Brain activity changes associated with treadmill training after stroke. *Stroke* 40, 2460-2467.

Fraser, A.M., and Swinney, H.L. (1986). Independent coordinates for strange attractors from mutual information. *Phys Rev A* 33, 1134-1140.

Frazzitta, G., Maestri, R., Uccellini, D., Bertotti, G., and Abelli, P. (2009). Rehabilitation treatment of gait in patients with Parkinson's disease with freezing: a comparison between two physical therapy protocols using visual and auditory cues with or without treadmill training. *Mov Disord* 24, 1139-1143.

Frenkel-Toledo, S., Giladi, N., Peretz, C., Herman, T., Gruendlinger, L., and Hausdorff, J.M. (2005). Treadmill walking as an external pacemaker to improve gait rhythm and stability in Parkinson's disease. *Mov Disord* 20, 1109-1114.

Hak, L., Houdijk, H., Steenbrink, F., Mert, A., Van Der Wurff, P., Beek, P.J., and Van Dieen, J.H. (2012). Speeding up or slowing down?: Gait adaptations to preserve gait stability in response to balance perturbations. *Gait Posture*.

Halsband, U., Ito, N., Tanji, J., and Freund, H.J. (1993). The role of premotor cortex and the supplementary motor area in the temporal control of movement in man. *Brain* 116 ( Pt 1), 243-266.

Hausdorff, J.M., Peng, C.K., Ladin, Z., Wei, J.Y., and Goldberger, A.L. (1995). Is walking a random walk? Evidence for long-range correlations in stride interval of human gait. *Journal of Applied Physiology* 78, 349-358.

Hausdorff, J.M., Purdon, P.L., Peng, C.K., Ladin, Z., Wei, J.Y., and Goldberger, A.L. (1996). Fractal dynamics of human gait: stability of long-range correlations in stride interval fluctuations. *Journal of Applied Physiology* 80, 1448-1457.

Hausdorff, J.M., Rios, D.A., and Edelberg, H.K. (2001). Gait variability and fall risk in community-living older adults: a 1-year prospective study. *Archives of physical medicine and rehabilitation* 82, 1050.

Herman, T., Giladi, N., Gurevich, T., and Hausdorff, J.M. (2005). Gait instability and fractal dynamics of older adults with a "cautious" gait: why do certain older adults walk fearfully? *Gait Posture* 21, 178-185.

Hof, A., Vermerris, S., and Gjaltema, W. (2010). Balance responses to lateral perturbations in human treadmill walking. *Journal of Experimental Biology* 213, 2655-2664.

Jordan, K., Challis, J.H., Cusumano, J.P., and Newell, K.M. (2009). Stability and the time-dependent structure of gait variability in walking and running. *Hum Mov Sci* 28, 113-128.

Kang, H.G., and Dingwell, J.B. (2006). Intra-session reliability of local dynamic stability of walking. *Gait Posture* 24, 386-390.

Kennel, M.B., Brown, R., and Abarbanel, H.D. (1992). Determining embedding dimension for phase-space reconstruction using a geometrical construction. *Phys Rev A* 45, 3403-3411.

Kim, H.S., Eykholt, R., and Salas, J. (1999). Nonlinear dynamics, delay times, and embedding windows. *Physica D: Nonlinear Phenomena* 127, 48-60.

Kuo, A.D. (2002). The relative roles of feedforward and feedback in the control of rhythmic movements. *Motor Control* 6, 129-145.

Lim, I., Van Wegen, E., De Goede, C., Deutekom, M., Nieuwboer, A., Willems, A., Jones, D., Rochester, L., and Kwakkel, G. (2005). Effects of external rhythmical cueing on gait in patients with Parkinson's disease: a systematic review. *Clin Rehabil* 19, 695-713.